\documentclass[amstex,aps,showpacs, twocolumn]{revtex4-2}
\usepackage{graphicx}
\usepackage{color}
\usepackage{bm}
\usepackage{longtable}
\usepackage{amsmath}
\usepackage{amsfonts}
\usepackage{amssymb}%
\usepackage{lmodern}

\begin{document}
\title{Leggett-Garg test of macrorealism using indefinite causal order of measurements}
\author{A. K. Pan}
\email{akp@phy.iith.ac.in}
	 \affiliation{Department of Physics, Indian Institute of Technology Hyderabad, Kandi, Telengana-502284, India }

\begin{abstract}
Macrorealism is a belief that constitutes the core of our perception of reality in the everyday world.  The Leggett-Garg (LG) test is a conceptually elegant approach for probing the compatibility between the notion of macrorealism and quantum theory. However, a conclusive LG test hinges on how one fixes the operational invasiveness loophole, i.e., how the statistical form of non-invasive measurability assumption is guaranteed in an LG test. Despite many attempts to close this loophole, no consensus has been achieved yet. In this work, we propose a simple and elegant scheme based on indefinite causal order in quantum switch experiment, which enables us to close this loophole, and eventually, the LG test becomes a conclusive test of macrorealism. 
\end{abstract}
\maketitle
The notion of macrorealism - the macroscopic objects in our everyday world has well-defined properties irrespective of being actually observed - is a well-accepted belief. It is the same tenet that constitutes the core of our classical worldview about the nature of reality.  The moon exists objectively and has a definite position in the sky even when no one looks.  However, since the inception of quantum theory, the extent to which this view is compatible with the standard framework has triggered intense debates among physicists and philosophers and remains the subject of active research. 

The pertinent question is how such a macrorealistic worldview fits into the quantum theory unless the division between micro and macro-world is imposed. Even so, where to put the `cut' (\emph{\ 'a la} Heisenberg \cite{He25}) and how the quantum theory fixes it remains a ticklish question. This issue first gets its firm ground through the Schr$\ddot{o}$ dinger's famous thought experiment \cite{sch} involving an unfortunate cat. From his viewpoint, there is no logical reason to describe a cat (macro-object) and a neutron (micro-object) in different levels of reality within quantum formalism, in contrast to Bohr's view of the `quantum system' and `classical apparatus'. 

There have been many attempts to pose the appropriate questions relevant to this issue. We may try here to put it in a somewhat different manner by first accepting the fact that we are undoubtedly \emph{not} interested in the question that in which limit quantum theory reduces to the classical one. Rather, we seek to answer whether a macro-object follows our everyday worldview of realism if the universality of quantum theory is accepted. It is desired that being an universal theory  the quantum theory must be able to recover the nature of reality involving macro-objects.  Many  proposals have been put forwarded to probe this issue, viz. the decoherence program \cite{zur}, coarse-graining measurements \cite{bruk}, unified micro-macro description \cite{ghi,bassi}, macro-objectification of quantum coherence of large objects \cite{arndt,gerlich}. However, they do not directly address the (in)compatibility between macrorealism and quantum theory. 

The Leggett and Garg (LG)  test \cite{leggett85,leggett02} is a conceptually elegant, direct and empirically testable approach. It consists of two main assumptions. i) \emph{Macrorealism }{\emph per se}: Macroscopic system remains in a definite state (maybe unknown) at any instant of time irrespective of observation.  ii) \emph{Non-invasive measurability}: State of the macro-system can be determined (at least in principle) without affecting the system itself and its subsequent dynamics. The first assumption seems quite obvious and fits with our perception of the nature around us while dealing with macro-object. The second one is also derivable from our everyday experience. LG test claims to establish the untenability of macrorealism in quantum theory, hence violating one or both assumptions. 

Let us briefly recapitulate the essence of LG framework. Consider the measurement of a dichotomic  observable $\hat{M}_{1}$ at three different times $t_1$, $t_2$ and $t_3$ $ (t_3 \geq t_2 \geq t_1 )$. In between the measurements at two different times, the system unitarily evolves under a suitable Hamiltonian $\hat{H}$, so that $M_{j}=U^{\dagger}_{t_j-t_i}M_{i}U_{t_j-t_i}$ where $i,j=1,2,3$ and $j>i$. Here $U_{t_j-t_i}=e^{-i \hat{H}(t_j-t_i)/\hbar}$. The three-time Leggett-Garg inequality (LGI)  can be written as 
\begin{equation}
\label{eq1}
K_{3}=1+\sum\limits_{m_{i},m_{j}=\pm1}m_i m_j \langle \hat{M_{i}} \hat{M_{j}}\rangle  \geq {0} 
\end{equation}
which is derived based on the above two assumptions. The correlation $\langle {M_{i}} {M_{j}}\rangle=\sum\limits_{m_{i},m_{j}=\pm1} m_{i} m_{j} p(m_{i}, m_{j})$  and $p(m_{i}, m_{j})$ is the pair-wise sequential joint probability of outcomes $m_{i}$ and  $m_{j}$. The quantum violation of LGI in Eq.(\ref{eq1}) for any two-level system is  extensively studied in the literature both theoretically  \cite{wvlgi,maroney,clemente15,clemente16,hali16aa,halli16,swati17,halli17,pan17,kumari18,bose,pan18,uola,halli19,halli19a} and experimentally \cite{robens,pala,dresexp,goggin,knee,avella,joar}. For an extensive review, we refer \cite{emary}.

We note that the standard three-time LGIs in Eq. (\ref{eq1}) do not provide the necessary and sufficient conditions for macrorealism. The quantum violation of  LGIs implies the violation of macrorealism, but the converse does not hold. The crucial question we address in this work is whether the violation of an LGI unequivocally warrants that the system is not behaving macrorealistically.  

 To pose this question more precisely, a comparison between LG and Bell tests would be helpful.  LGIs are often considered as the temporal analog of Bell's inequalities primarily due to the structural resemblance between Bell-CHSH inequalities and four-time LGIs.   However, LG test shows serious pathology \cite{wvlgi,clemente15,clemente16,halli16,swati17,halli17,pan17,kumari18,pan18,uola,halli19,halli19a} in comparison to a Bell test of local realism. In the latter case, the statistical version of locality condition, the no-signaling in space is satisfied in any physical theory. Hence, the Bell test unequivocally establishes the violation of both locality and realism. On the other hand, the operational non-invasiveness (commonly known as no-signaling in time) - the statistical version of the non-invasive measurability assumption is not generally satisfied in quantum theory. A skeptic (stubborn macrorealist) can then rightfully argue that she can salvage macrorealism in quantum theory by simply abandoning non-invasive measurability. Thus, unless the operational non-invasiveness loophole (also known as clumsiness loophole \cite{wild}) is fixed in the experiment, the LG test cannot be considered as a conclusive test of macrorealism. 

Many attempts have already been made to close this loophole by proposing various forms of measurement schemes. Leggett himself was well aware of it, and he proposed the ideal negative-result measurement. Many experimental tests  \cite{knee,pala,goggin,robens,joar} of LGI have been performed by using it. However, since in negative-result measurement the collapse of state occurs, i.e., system and its subsequent dynamics \cite{maroney} is disturbed, then up to what extent this approach satisfies the operational non-invasiveness remains debatable. Another approach is by using the technique of weak measurement \cite{aav} where the strength of the measurement is possible to adjust and, in principle, the back action (the invasiveness) on the system can be reduced to an arbitrarily small amount \cite{dresexp,avella}. Again, how this idea of assuming weak coupling (still cause small back-action) in quantum measurement will convince a macrorealist remains unsatisfactory \cite{maroney,halli16,hali16aa,pan20}.  

LG test involves the measurement of non-commuting observables. There is no unique prescription to fix the joint probability for such observables in the standard framework of quantum measurement theory. However, it is common in the literature to use quasiprobability in such a scenario. In the LG scenario, one may also define suitable quasiprobability as pointed out by Halliwell  \cite{halli16}. We demonstrate how suitably defined quasiprobability enables a conclusive LG test by providing the explicit measurement scheme based on the quantum switch experiment that uses indefinite causal order of measurements. We note here that the sequential correlation of two non-commuting observables in Eq. (\ref{eq1}) can also be written in a form 
\begin{eqnarray}
\label{mimj}
\langle {M_{i}} {M_{j}}\rangle_{seq}=\frac{1}{2}Tr\left[\{M_i, M_j\}\rho\right]
\end{eqnarray}
where $\rho$ is the density matrix. The proof of Eq. (\ref{mimj}) is given in Appendix A. This indicates a superposition of measurements of $M_{i}$ and $M_{j}$ are needed, which can be achieved through the indefinite order of measurements. We also argue here how our approach fixes the operational non-invasiveness loophole. 
  
A couple of remarks on the non-invasive measurability condition could be useful here. This condition can be interpreted in two different ways, contingent upon how the sequential measurements are performed in an experiment. Commonly three sequential correlations given in Eq. (\ref{eq1}) are performed involving the measurements at two times. This is analogous to the Bell test, and hence for a conclusive LG test, the operational non-invasiveness must be satisfied for the pair-wise case instead of the triple-wise case. This is a weaker reading of non-invasive measurability condition as pointed out in \cite{halli19}. The stronger version of the non-invasive measurability condition assumes the satisfaction of operational non-invasiveness for the sequential measurements at three times. We show that for both weaker and stronger readings of the non-invasive measurability condition, our scheme works. 

However, for better understanding of our idea, let us first present the two-time LG scenario involving two-time sequential correlation only. LGIs for two-time measurement scenario can be formulated as
\begin{eqnarray}
\label{lgq}
G_{2}=1+m_i \langle\hat{M_{i}}\rangle +  m_j\langle\hat{M_{j}}\rangle+ m_i m_j \langle\hat{M_{i}}\hat{M_{j}}\rangle \geq 0 
\end{eqnarray}
with $i,j=1,2,3$ and $j>i$. For different values of $m_i=\pm1$ and $m_j=\pm1$ we have four two-time LGIs. Halliwell \cite{halli16,halli17} argued that the above set of LGIs provides necessary and sufficient conditions for macrorealism in the two-time LG scenario.  

We shortly see that $G_{2}$ are actually quasiprobabilities. As usual, a quasiprobability follows all of Kolmogorov's axioms for probability except the property of positivity. Negative and complex quasiprobabilities have immense implications in quantum information processing \cite{ferrie1}, quantum tomography \cite{lundeen12} and quantum metrology \cite{nature}. It has also been demonstrated that the negativity and contextuality are equivalent notions of nonclassicality \cite{spek08}. Dressel \cite{dressel} discussed that there is a close connection between quasiprobability and weak value\cite{aav}. The widely  known quasiprobaility is the Wigner distribution \cite{wigner}. For discrete system,  one of the well-known quasiprobabilities  is Kirkwood distribution \cite{kirk} which is in general complex and the real part of of it is known as Merganau-Hill distribution \cite{mg,johan}. Given a density matrix $\rho$, Merganau-Hill distribution can be written as 
\begin{eqnarray}
q_{Q}(m_{i},m_{j})=Re[\langle m_{i}|m_{j}\rangle\langle m_{j}|\rho|m_{i}\rangle]
\end{eqnarray}
satisfying $\sum_{m_{i},m_{j}} q_{Q}(m_{i},m_{j})=1$.  Here $\{|m_{i}\rangle\}$ and $\{|m_{j}\rangle\}$ are two arbitrary bases on the same Hilbert space in which $\rho$ belongs. Using the moment expansion corresponding to dichotomic observables here we can write the quasiprobability in a macrorealistic model as  
\begin{eqnarray}
	q(m_i, m_j)= \frac{1}{4}(1+ m_i \langle M_{i}\rangle +m_j \langle M_{j}\rangle+m_i m_j\langle M_{i} M_{j}\rangle)
\end{eqnarray}
which is nothing but the two-time LGIs given by Eq.(\ref{lgq}), with a multiplicative factor 4. Thus, the negativity of a quasiprobability, i.e., $q_{Q}(m_i , m_j)<0$ implies the quantum violation of two-time LGIs.

Let us discuss the interesting properties of $q_{Q}(m_i, m_j)$. First, it correctly reproduces the sequential correlation in quantum theory, so that   
\begin{eqnarray}
\label{pcor}
\nonumber
	\langle M_{i} M_{j}\rangle=&\sum\limits_{m_i, m_j}m_i m_j \ q_{Q}(m_i ,m_j)\\
	&=\sum\limits_{m_i, m_j}m_i m_j  \ p(m_i, m_j)
\end{eqnarray}

Second, the most important property for the present work, $q_{Q}(m_i ,m_j)$ satisfies the operational non-invasivness in quantum theory, i.e., 

\begin{subequations}
\begin{eqnarray}
\label{oi1}
	p(m_j)=\sum\limits_{m_i}	q_{Q}(m_i, m_j)= Tr[|m_j\rangle\langle m_{j}|\rho] \\
	\label{oi2}
	p(m_i)=\sum\limits_{m_j}	q_{Q}(m_i, m_j) = Tr[|m_i\rangle\langle m_{i}|\rho] 
\end{eqnarray}
\end{subequations}
Explicitly, the Eqs. (\ref{oi1})  implies that $p(m_j)$ remains independent of the fact if a prior measurement of $M_i$ is performed. Hence the operational non-invasiveness condition is satisfied by the quasiprobabilities $q_{Q}(m_2 ,m_3)$. But $p(m_j)\neq \sum\limits_{m_i} p(m_i, m_j)$ in general. Since the correlations $\langle M_{i} M_{j}\rangle$ remains same one may then use  $q_{Q}(m_2 ,m_3)$ instead of  $p(m_2 ,m_3)$ to obtain the sequential correlations in three-time LG scenario.

Importantly, since operational non-disturbance remains satisfied, $q_{Q}(m_i , m_j)<0$ constitutes a conclusive test macrorealism provided it can be directly tested in the experiment. We show that the quantum switch experiment \cite{chi1,gos,guo,ban} realizing indefinite causal order of measurements can provide a direct test of $q_{Q}(m_i, m_j)$ and constitute a loophole-free test of macrorealism.

It is worthy of mentioning here that in an earlier work \cite{pan20} we have shown that a Mach-Zehender interference experiment itself enables a direct test of macrorealism. In that, no apparatus was involved in a standard path-only interference experiment, and the system itself serves as apparatus so that the measurement effectively causes no disturbance.  It was shown \cite{pan20} that the destructive interference, anomalous weak value, and the violation of two-time LGIs are well connected. The work in \cite{pan20} was further extended for double-slit interference experiment \cite{hali21} and several other interesting effects are discussed. It is then obvious but interesting here to note that there is a close connection between $q_{Q}(m_i ,m_j)$ and weak value. However, this weak value has no bearing on the weak coupling requirement in traditional weak measurement \cite{aav}. For pure state, $\rho=|\psi_{i}\rangle\langle \psi_{i}|$, we can write 
\begin{eqnarray}
q_{Q}(m_i, m_j)=(|m_{i}\rangle\langle m_{i}|)_{w}|\langle \psi_{i}|m_{j}\rangle|^{2}
\end{eqnarray}
 where $(|m_{i}\rangle\langle m_{i}|)_{w}$ is the weak value of the projector defined as $(|m_{i}\rangle\langle m_{i}|)_{w}=\langle m_{j}|m_{i}\rangle\langle m_{i}|\psi_{i}\rangle/\langle m_{j}|\psi_{i}\rangle$.  From Eq. (8) we can then say that when $q_{Q}(m_i, m_j)<0$, the weak value $(|m_{i}\rangle\langle m_{i}|)_{w}$  of the projector is anomalous and vice-versa.  There exists many choices of $|m_{i}\rangle$, $|m_{j}\rangle$ and $|\psi_{i}\rangle$ are available for which $q_{Q}(m_i, m_j)<0$ can be found.

The experimental determination of $q_{Q}(m_i, m_j)$  has been attempted by assembling the separately measured single and pair-wise correlations in Eq. (\ref{lgq})through  the  noninvasive continuous-in-time velocity measurements \cite{hali16aa,maji}. However, while measuring joint correlation of the non-commuting observable $M_{i}$ and $M_{j}$, the disturbance will play a role again.  We provide a scheme for a direct test of $q_{Q}(m_i, m_j)$ in quantum switch experiment \cite{chi1,gos,guo}.  which involves an interesting interferometric setup involving indefinite causal order of two non-commuting observables.  For this, let us note that for a pure state $\rho_{s}=|\psi_{i}\rangle\langle \psi_{i}|$, the quasipropabilities $q(m_i, m_j)$ can be written as 
\begin{align}
\label{anti}
	q_{Q}(m_i ,m_j) =\frac{1}{2}\langle\psi_{i}|\left\{\pi_{m_i},\pi_{m_j} \right\}|\psi_{i}\rangle
\end{align}
where $\pi(m_{i(j)})$ are the projectors $\pi(m_{i(j)})= |m_{i(j)}\rangle\langle m_{i(j)}|=(\mathbb{I}+m_{i(j)}M_{i(j)})/2$.

We consider an extended Mach-Zehender (MZ) setup (Fig.1) which realizes the indefinite causal order of observables called quantum switch \cite{chi1} which has recently been realized in experiment \cite{gos,guo}. The setup consists of a polarizing beam-splitter (PBS), a $50:50$ beam splitter $BS$ , eight  mirrors (not marked), a phase-shifter $(PS)$ and two detectors ($D_{1}$ and $D_{2}$). The photons having initial state  $|\psi\rangle\otimes|+\rangle$ where $|\psi\rangle$ and $|+\rangle=\left(|H\rangle + |V\rangle\right)/\sqrt{2}$ are path and system states respectively. After the PBS the state becomes \begin{equation}
|\Psi\rangle=\frac{|\psi_{H}\rangle|H\rangle +|\psi_{V}\rangle|V\rangle}{\sqrt{2}} 
\end{equation}

\begin{figure}[ht]
\includegraphics[width=0.55\textwidth]{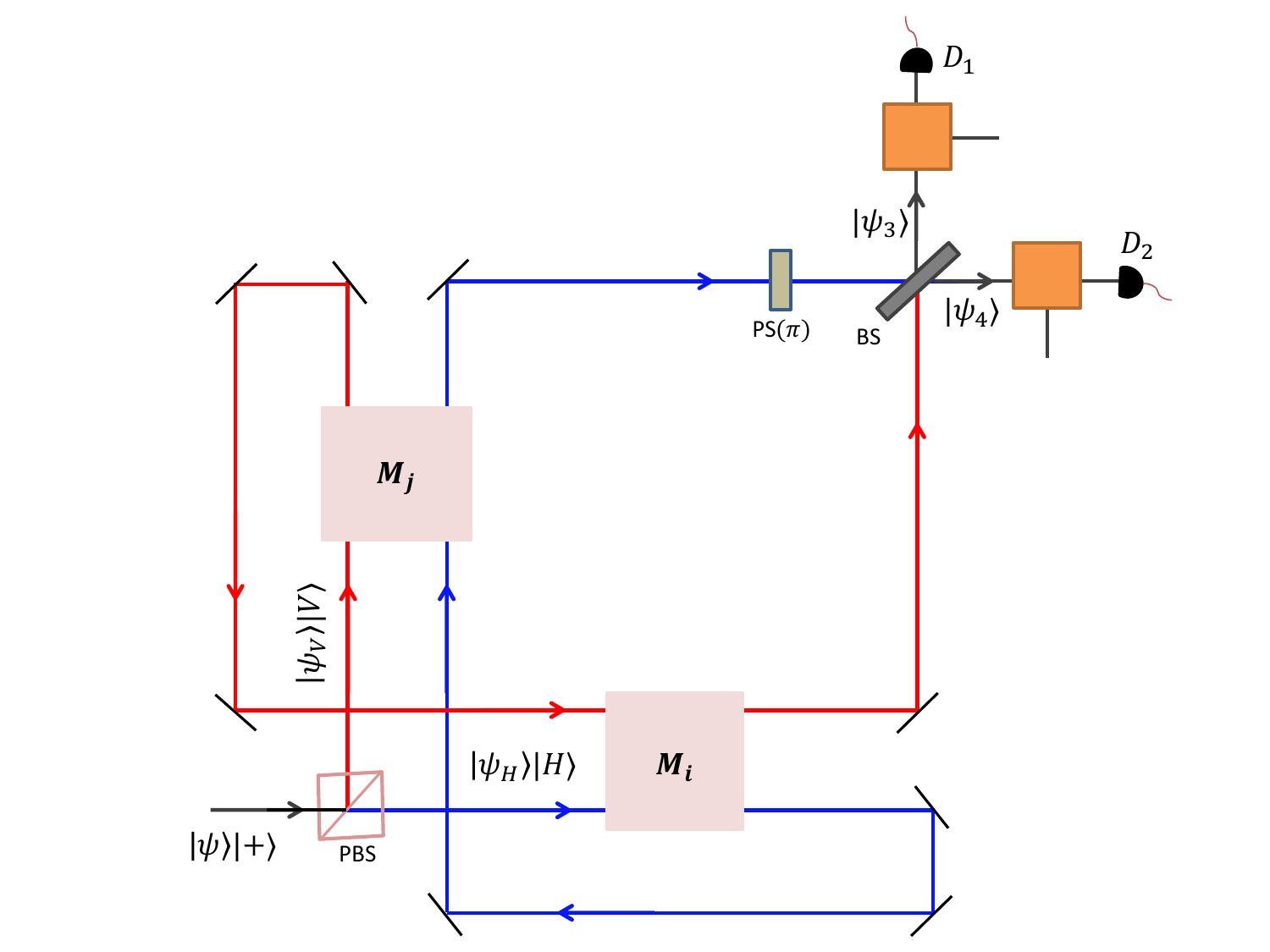}
\vskip -0.3cm
\caption{The quantum SWITCH experiment realizing indefinite causal order of measurements. See text for details.}
\end{figure}
After the PBS, the photon passes through the unitary channels performing the measurements of  $M_{i}$ and $M_{j}$ but indefinitely in causal order. This action is the key to obtain the anti-commutation in Eq. (\ref{anti}). For our purpose here, we consider a particular outcomes $m_{i(j)}$ of $M_{i(j)}$, i.e., we realize the statistics corresponding to projectors $\pi_{m_{i(j)}}$.

After the action of phase-shifter (PS) along the channel $|\psi_{H}\rangle$ and recombination at BS followed by polarization measurement in $\{+,-\}$ basis, the state of the photon becomes
\begin{eqnarray}
\label{finalpsi}
\nonumber
	|\Psi_{F}\rangle&=&\frac{1}{2\sqrt{2}}\Big[\Big(\{\pi_{m_i}, \pi_{m_j}\} |+\rangle +[\pi_{m_i}, \pi_{m_j}] |-\rangle\Big)|\psi_{3}\rangle\\
		&+& i\Big([\pi_{m_j}, \pi_{m_i}] |+\rangle -\{\pi_{m_i}, \pi_{m_j}\} |-\rangle\Big)|\psi_{4}\rangle\Big]
\end{eqnarray}
where  we have written the polarization in $\{|+\rangle, |-\rangle\}$ basis.  
Further, the post-selection in the state $|+\rangle$ along the path $|\psi_{3}\rangle$ or in the state $|-\rangle$ along the path $|\psi_{4}\rangle$ provide the quasiprobability in Eq. (\ref{anti}), if $|\psi_{i}\rangle$ is identified as $|+\rangle$. The detailed calculation of Eq. (\ref{finalpsi}) is given in Appendix B.  Hence, the two-time LGI can be experimentally tested while the operational non-invasiveness is satisfied. 

Using the $q_{Q}(m_i, m_j)$ the pair-wise correlations $\langle M_{i} M_{j}\rangle$ can also be obtained. The three-time LGI $K_{3}$ in Eq. (\ref{eq1}) can thus be tested when only pair-wise correlations are considered. Hence, the quasiprobility $q_{Q}(m_i, m_j)$ enables the loophole-free test of two-time LGI.

We now consider the case of three-time LGI involving triple-wise correlations. For this, we introduce doubly Kirkwood quasiprobability distribution as 
\begin{equation}
q_{Q}(m_{1},m_{2},m_{3})=\frac{1}{2}Re\left[\langle m_{1}|m_{2}\rangle \langle m_{2}| \rho|m_{3}\rangle\langle m_{3}|m_{1}\rangle\right] 
\end{equation}
where $\{m_{1}\}$, $\{m_{2}\}$, $\{m_{3}\}$ are three different bases on the same Hilbert space in which $\rho$ belongs. We can identify the bases corresponding to the measurements of $M_{1}$, $M_{2}$ and $M_{3}$ in LG scenario. For the pure state $\rho=|\psi\rangle\langle\psi|$, the quasiprobability $q_{Q}(m_{1},m_{2},m_{3})$ can be written as 
\begin{eqnarray}
q_{Q}(m_{1},m_{2},m_{3})=\frac{1}{2}\frac{Re\left[q_{Q}(m_{1},m_{3}) q^{\ast}_{Q}(m_{2},m_{3})\right]}{|\langle \psi|m_{3}\rangle|^{2}} 
\end{eqnarray}
It is straightforward to check that the sequential correlation remains same as
\begin{eqnarray}
\nonumber
	\langle M_{j} M_{k}\rangle=&\sum\limits_{m_i, m_j,m_k}m_j m_k \ q_{Q}(m_i ,m_j,m_{k})\\
	&=\sum\limits_{m_j, m_k}m_j m_k  \ p(m_j, m_k)
\end{eqnarray}
where $j,k=1,2,3$ with $j<k$. Another important point property $q_{Q}(m_{i},m_{j},m_{k})$ is that it satisfies all the operational non-invasiveness conditions so that
\begin{eqnarray}
p(m_{3})=\sum\limits_{m_1,m_2} q_{Q}(m_{1},m_{2},m_{3})=Tr[|m_{3}\rangle\langle m_{3}|\rho]\\
p(m_{2})=\sum\limits_{m_1,m_3} q_{Q}(m_{1},m_{2},m_{3})=Tr[|m_{2}\rangle\langle m_{2}|\rho]
\end{eqnarray}
and 
\begin{eqnarray}
p(m_{1}, m_{2})&=&\sum\limits_{m_2} q_{Q}(m_{1},m_{2},m_{3})\\
\nonumber
&=&Tr[|m_{1}\rangle\langle m_{1}|\rho|m_{1}\rangle\langle m_{1} |m_{3}\rangle\langle m_{3}|]\\
p(m_{2},m_{3})&=&\sum\limits_{m_1} q_{Q}(m_{1},m_{2},m_{3})\\
\nonumber
&=&Tr[|m_{2}\rangle\langle m_{2}|\rho|m_{2}\rangle\langle m_{2} |m_{3}\rangle\langle m_{3}|]
\end{eqnarray}
Note that $q_{Q}(m_{i},m_{j},m_{k})>0$ does not imply three-time LGI in Eq. (\ref{eq1}) as was the case for two-time LGIs. In recent work, Majdy et al. \cite{maj} proposed a three-time LGI similar to the two-time LGI as

\begin{eqnarray}
\label{g3}
G_{3}&=&\frac{1}{8}\big(1+3\sum\limits_{i=1}^{3} m_i \langle M_{i}\rangle \\
\nonumber
&+& \sum\limits_{i\neq j=1}^{3}m_i m_j\langle M_{i} M_{j}\rangle+m_1 m_2 m_{3}\langle M_{1} M_{2} M_{3}\rangle\big)
\end{eqnarray}  
This inequality is violated in quantum theory even when two-time and three-time LGIs in Eqs. (\ref{eq1}-\ref{lgq}) are not violated. Importantly, $q_{Q}(m_{i},m_{j},m_{k})>0$ implies the inequality  Eq. (\ref{g3}) with a multiplicative factor $1/8$. Hence, a direct test of  $q_{Q}(m_{i},m_{j},m_{k})$ provides a conclusive test of three-time LGIs in Eq. (\ref{g3}). A simple modification of the setup presented in Fig. 1 will serve this purpose. Note here that both weaker and stronger versions of operational non-invasiveness are satisfied here.

One can also test the standard three-time LGI given by Eq. (\ref{eq1}) by considering a suitable combination of two three-time quasiprobabilities as
\begin{equation}
q_{Q}(m_{i},-m_{j},m_{k})+ q_{Q}(-m_{i},m_{j},-m_{k})\geq 0 
\end{equation}
This means that the negativity of both of them provides the quantum violation of three-time LGI, but the negativity of one of them does not guarantee it. 
 For testing the LGIs in Eq. (\ref{eq1}), quasiprobabilities $q_{Q}(m_{i},-m_{j},m_{k})$ and $ q_{Q}(-m_{i},m_{j},-m_{k})$ has to be separately measured. This can also be done by incorporating simple modifications in the setup in Fig.1.  We have not provided a detailed description of such modification, but it is quite obvious. 

In conclusion, we have provided a protocol to demonstrate the loophole-free LG test of macrorealism through the quantum switch experiment that realizes the indefinite causal order of measurements. Two-time LGIs in Eq. (\ref{lgq}) three-time LGIs provide the necessary and sufficient condition for macrorealism \cite{halli16}, but the same is not valid for three-time LGIs \cite{clemente16}. Note that except for the CHSH inequality, the Bell test does not provide the necessary and sufficient condition for local realism. However, the violation of a Bell's inequality unequivocally warrants the violation of local realism. However, the same argument does not hold for the LG test of macrorealism, as the violation of LGIs can be attributed to the violation of operational non-invasiveness alone. A skeptic can then always salvage the macrorealism \emph{per se}.  Hence, in contrast to a Bell violation (where no-signaling in spatially separated systems is always satisfied), the LG violation for macrorealism is inconclusive unless the operational non-invasiveness loophole is fixed in the experiment. 

We stress again that in this work, we are not interested in whether LGIs provide necessary and sufficient conditions. Instead, we are concerned whether the LG test is conclusive enough to refute the LG notion of macrorealism in quantum theory whenever we have the quantum violation of a given LGI. We demonstrated that suitably defined quasiprobabilities - the ones that reproduce the sequential correlations correctly and satisfy the operational invasiveness- can provide a conclusive LG test of macrorealism while negative. We demonstrated that such quasiprobabilities could be directly tested in a quantum switch experiment that realizes the indefinite causal order of measurements.   Our measurement scheme thus closes the operational non-invasiveness loophole so that the LG test becomes a conclusive test of macrorealism when LGI is violated by quantum theory. This would then convince a skeptic that macrorealism cannot be salvaged by abandoning the non-invasive measurability condition and, indeed, incompatible with quantum theory. Our scheme thus provides a decisive and conclusive test of macrorealism that the LG test alluded to.

We may remark here that an alternate formulation for testing macrorealism - the no-signaling in time conditions that provides necessary and sufficient conditions for macrorealism for a three-time LG scenario. We agree with Halliwell \cite{halli16} that NSIT is more of a test of non-invasive measurability than macrorealism \emph{per se}. On the other hand, LGIs test both the assumptions provided the operational non-invasiveness is guaranteed in the experiment. Also, there remains debate regarding the relationship between the two assumptions involved in the LG framework. As per Leggett's viewpoint, the non-invasive measurability implies macrorealism \emph{per se} and this view is supported in \cite{clemente15}. However, Maroney and Timpson \cite{maroney} argued that non-invasive measurability condition is independent of macrorealism \emph{per se}. Such kind of independence is actually desired to test LGIs in a theory-independent framework.from the viewpoint of the ontological model framework advocated by Spekkens \cite{spek05}  However, this issue is not yet settled. Studies along this direction could be an exciting avenue for future works.  
\appendix
\section{Derivation of Eq. (\ref{mimj})}
The sequential correlation in Eq. (\ref{mimj})  can be calculated as follows. Given a density matrix $\rho$ the correlation function for the sequential measurement of two observables $\hat{M}_i$ and $\hat{M}_j$ can be written as
\begin{eqnarray}
\label{s2}
\langle \hat{M}_{i}\hat{M}_{j}\rangle_{seq}=\sum_{m_{i},m_{j}=\pm1}{m_{i} m_{j}}P({m_{i}},{m_{j}})
\end{eqnarray}
with $j>i$ and the joint probability of the sequential quantum measurement is  given by 
\begin{eqnarray}
\label{s22}
P({m_i},{m_j})= Tr\left[\left(\sqrt{E_{M_i}^{m_i}}^{\dagger}\rho \sqrt{E_{M_i}^{m_i}} \right)\Pi_{M_j}^{m_j}\right]
\end{eqnarray}
where $E_{M_i}^{m_i}$ are the positive-operator-valued-measures (POVMs) corresponding to the measurement of a dichotomic $\hat{M}_i$ with $m_i=\pm 1$. In a sequential measurement, we may consider the measurement of $\hat{M}_j$ is a projective measurement. If $\{\Pi_{M_j}^{m_j}\}$ are the projectors of dichotomic observable $\hat{M}_j$ then $\hat{M}_j=\Pi_{M_j}^{+}-\Pi_{M_j}^{-}$ with $m_j=\pm 1$. 

By using Eq. (\ref{s22}) we can re-write Eq.(\ref{s2}) as,
\begin{eqnarray}
\label{s3}
\nonumber
&&\langle \hat{M}_i\hat{M}_j\rangle_{seq}=\sum_{m_i=\pm1}{m_i}Tr\left[\left(\sqrt{E_{M_i}^{m_i}}^{\dagger}\rho \sqrt{E_{M_i}^{m_i}}\right) \Pi_{M_j}^{+}\right]\\
&-&\sum_{m_i=\pm1}{m_i}Tr\left[\left(\sqrt{E_{M_i}^{m_i}}^{\dagger}\rho \sqrt{E_{M_i}^{m_i}} \right) \Pi_{M_j}^{-}\right]
\end{eqnarray}
Let us consider the POVMs corresponding to the measurement of $M_{i}$ as 
\begin{align}
\label{em1}
	E_{M_i}^{\pm}= \frac{1\pm\lambda}{2}\Pi_{M_i}^{\pm} + \frac{1\mp\lambda}{2} \Pi_{M_i}^{\mp}
\end{align}
where $\lambda$ is the unsharpness parameter and  $\Pi_{M_i}^{\pm}$ are projectors of $\hat{M}_i$. 
Plugging Eq. (\ref{em1}) in Eq. (\ref{s3}), we get 
\begin{eqnarray}
\label{s5}
\nonumber
\langle \hat{M}_i\hat{M}_j\rangle_{seq}= \lambda \Big(Tr[\rho\left(\Pi_{M_i}^{+}{M_j} \ \Pi_{M_i}^{+} -\Pi_{M_i}^{-} {M_j} \ \Pi_{M_i}^{-}\right)]\Big)\\
\end{eqnarray} 
Putting $\Pi_{M_i}^{\pm}=(\mathbb{I}\pm\hat{M}_i)/2$ for dichotomic observable,  Eq.(\ref{s5}) can be simplified as 
\begin{eqnarray}
\label{s6}
\langle\hat{M}_i\hat{M}_j\rangle_{seq}=\frac{\lambda}{2}Tr\left[\rho \left\{\hat{M}_i,\hat{M}_j\right\}\right] 
\end{eqnarray} 
which is the Eq. (\ref{mimj}) in the main  text. If the measurement of $M_i$ is sharp we need to take $\lambda=1$.  \section{Explicit derivation of Eq. (\ref{finalpsi})}
As considered in the main text let us consider the system state $|+\rangle=\frac{1}{\sqrt{2}} \left(|H\rangle +|V\rangle\right)$ entering the MZ setup and $|\psi_{H}\rangle$ and $|\psi_{V}\rangle$ are the control states so that the application of indefinite causal order of unitaries can be written as 
\begin{align}
\label{uij}
	\mathcal{U}_{ij}=U_{i}U_{j}\otimes |\psi_{H}\rangle\langle \psi_{H}| +U_{j}U_{i}\otimes |\psi_{V}\rangle\langle \psi_{V}|
\end{align}
where $U_{i}$ and $U_{j}$ are the unitary operators corresponding to the measurements of $M_{i}$ and $M_{j}$ respectively. The state $\rho=|\Psi\rangle\langle \Psi|$ after PBS is given by Eq. (10) in the manuscript. The state $\rho$ passes through $\mathcal{U}_{ij}$.  We consider the initial apparatus states are $|\xi_{i}\rangle$ and $|\xi_{i}\rangle$  corresponding to the measurements of $M_{i}$ and $M_{j}$ respectively. Hence the total system-control-apparatus state can be written as $\rho_{SA}=\rho \otimes |\xi_{i}\rangle\langle \xi_{i}| \otimes |\xi_{j}\rangle\langle \xi_{j}|$. After the unitary operation as in Eq. (\ref{uij}) the state $\rho_{SA}$ becomes

\begin{align}
	\rho^{\prime}_{SA}=\mathcal{U}_{ij} \ \rho_{SA} \ \mathcal{U}_{ij}^{\dagger}
\end{align}
 which can be explicitly written by using Eq. (\ref{uij}) as 
\begin{eqnarray}
\nonumber
	\rho^{\prime}_{SA}&=&\frac{1}{2}\Big[U_{i} U_{j}|\xi_{i}\rangle\langle \xi_{i}| \otimes |\xi_{j}\rangle\langle \xi_{j}|\otimes |H\rangle\langle H|U_{i}^{\dagger} U_{j}^{\dagger}\otimes |\psi_{H}\rangle\langle \psi_{H}|\\
	\nonumber
	&+&\left(U_{i} U_{j}|\xi_{i}\rangle\langle \xi_{i}| \otimes |\xi_{j}\rangle\langle \xi_{j}|\otimes |H\rangle\langle V|U_{j}^{\dagger} U_{i}^{\dagger}\otimes |\psi_{H}\rangle\langle \psi_{V}|\right)\\
	\nonumber
	&+&\left(U_{j} U_{i}|\xi_{i}\rangle\langle \xi_{i}| \otimes |\xi_{j}\rangle\langle \xi_{j}|\otimes |V\rangle\langle H|U_{i}^{\dagger} U_{j}^{\dagger}\otimes |\psi_{V}\rangle\langle \psi_{H}|\right)\\
	&+&U_{j} U_{i}|\xi_{i}\rangle\langle \xi_{i}| \otimes |\xi_{j}\rangle\langle \xi_{j}|\otimes |V\rangle\langle V|U_{j}^{\dagger} U_{i}^{\dagger}\otimes |\psi_{V}\rangle\langle \psi_{V}|\Big]
\end{eqnarray}
 
Now, by considering the post-measurement apparatus states are $|\xi(m_{i})\rangle$ and $|\xi(m_{j})\rangle$ for the measurements of $M_{i}$ and $M_{j}$ respectively, and by following the standard notion of generalized quantum measurement theory we can define the measurement operators  as $N_{m_{i}}=\langle \xi(m_{i})| U_{i}|\xi_{i}\rangle$ and $N_{{m_{j}}}=\langle \xi(m_{j})| U_{j}|\xi_{j}\rangle$. The POVMs can be written as $E_{M_{i}}^{m_{i}}=N_{m_{i}}^{\dagger}N_{m_{i}}$ and $E_{M_{j}}^{m_{j}}=N_{m_{j}}^{\dagger}N_{m_{j}}$. For particular outcomes of $m_{i}$ and $m_{j}$, we can write the reduced system-control state as 
\begin{eqnarray}
	\rho^{\prime}_{S}&=&\frac{1}{2}\Big[N_{i} N_{j}|H\rangle\langle H| N_{i}^{\dagger} N_{j}^{\dagger}\otimes |\psi_{H}\rangle\langle \psi_{H}|\\
	\nonumber
	&+&N_{i} N_{j}|H\rangle\langle V|N_{j}^{\dagger} N_{i}^{\dagger}\otimes |\psi_{H}\rangle\langle \psi_{V}|\\
	\nonumber
	&+&N_{j} N_{i} |V\rangle\langle H|N_{i}^{\dagger} N_{j}^{\dagger}\otimes |\psi_{V}\rangle\langle \psi_{H}|\\
	\nonumber
	&+&N_{j} N_{i} |V\rangle\langle V|N_{j}^{\dagger} N_{i}^{\dagger}\otimes |\psi_{V}\rangle\langle \psi_{V}|\Big]
\end{eqnarray}
which can also be written as $\rho^{\prime}_{S}=|\Psi^{\prime}\rangle\langle \Psi^{\prime}|$ where 
\begin{align}
	|\Psi^{\prime}\rangle=\frac{1}{\sqrt{2}}\left((N_{i} N_{j} |\psi_{H}\rangle|H\rangle + (N_{j} N_{i} |\psi_{V}\rangle|V\rangle \right)
\end{align}
For projective measurements we can take $N_{m_i}$ and $N_{m_j}$ as $\pi_{m_{i}}$ and $\pi_{m_{j}}$ respectively. Now, after the application of a phase-shifter (introducing $\pi$ phase shift) along $|\psi_{H}\rangle$ and the  recombination  the state at second BS followed by a polarization measurement,  we get
\begin{eqnarray}
\label{psif}
	\nonumber
	|\Psi_{F}\rangle&=&\frac{1}{2\sqrt{2}}\Big[\Big(\{\pi_{m_i}, \pi_{m_j}\} |+\rangle +[\pi_{m_i}, \pi_{m_j}] |-\rangle\Big)|\psi_{3}\rangle\\
	&+& i\Big([\pi_{m_j}, \pi_{m_i}] |+\rangle -\{\pi_{m_i}, \pi_{m_j}\} |-\rangle\Big)|\psi_{4}\rangle\Big]
\end{eqnarray}
which is the Eq. (\ref{finalpsi}) in the main text. To get Eq. (\ref{psif}), we take $|\psi_{H}\rangle=\frac{1}{\sqrt{2}}\left(|\psi_{3}\rangle+i|\psi_{4}\rangle\right)$ and $|\psi_{V}\rangle=\frac{1}{\sqrt{2}}\left(i|\psi_{3}\rangle+|\psi_{4}\rangle\right)$ for the BS  along with $|H\rangle=\frac{1}{\sqrt{2}}\left(|+\rangle+|-\rangle\right)$ and $|V\rangle=\frac{1}{\sqrt{2}}\left(|+\rangle-|-\rangle\right)$ for polarization measurement, as shown in Fig. 1.

Finally, if the post-selection is performed in the state $|\psi_{3}\rangle |+\rangle$, we get the quasiprobability $q_{Q}(m_i, m_j)=\frac{1}{2}\langle +|\{\pi_{m_i}, \pi_{m_j}\}|+\rangle$ with a multiplicative factor $1/\sqrt{2}$.

\acknowledgments
 The author is indebted to Prof. J. J. Halliwell for many stimulating email exchanges on LGIs over the years and acknowledges the support from the research grant DST/ICPS/QuEST/2019/4. 

\end{document}